\documentclass[aps,prd,groupedaddress,showpacs]{revtex4}

\usepackage{amsfonts}
\usepackage{graphicx}

\newtheorem{theorem}{Theorem}
\newtheorem{corollary}{Corollary}
\def\endprf{\hfill  {\vrule height6pt width6pt depth0pt}\medskip}
\newenvironment{proof}{\noindent {\bf Proof} }{\endprf\par}

\begin{document}


\title{Infrared Gluon and Ghost Propagators}


\author{Marco Frasca}
\email[]{marcofrasca@mclink.it}
\affiliation{Via Erasmo Gattamelata, 3 \\ 00176 Roma (Italy)}


\date{\today}

\begin{abstract}
We derive the form of the infrared gluon propagator by proving a mapping in the infrared of 
the quantum Yang-Mills  and $\lambda\phi^4$ theories. The equivalence is complete at a classical level. But while at a quantum level, the correspondence is spoiled by quantum fluctuations in the ultraviolet limit,
we prove that it holds in the infrared where the coupling constant happens to be very large. 
The infrared propagator is then obtained from the quantum field theory of the scalar field producing a full spectrum. 
The results are in fully agreement with recent lattice computations. We get a finite propagator at zero momentum, 
the ghost propagator going to infinity as $1/p^{2+2\kappa}$ with $\kappa=0$.
\end{abstract}


\maketitle

\section{Introduction}

Infrared behavior of the gluon and ghost propagators appears at odds with theoretical computations done
with Dyson-Schwinger (DS) equations with a given truncation scheme \cite{as,cf}. This approach foresees a
gluon propagator going to zero and a ghost propagator going to infinity faster than $1/p^2$ in the infrared.
At the same time, the running coupling is expected to reach a fixed point. Analysis done on a compact
manifold gives a contradictory result with a gluon propagator bending toward zero but a running
coupling going to zero rather to a fixed a point \cite{cf2}. The authors in this latter work claim
that presently Yang-Mills on the lattice has not yet reached the right distances (10-15 fm) to
appreciate the bending of the gluon propagator toward zero.

Recent lattice results given at the Lattice 2007 Conference extended the analysis to the required
limits \cite{ilg1,ilg2,cuc,ste}. Again, none of the above behaviors are seen in both gluon and ghost propagators making no more believable the DS equations approach, in the way it has been devised, as a viable 
description of Yang-Mills theory in the infrared. These lattice results depict a different scenario.
They show that the gluon propagator goes to a constant at a diminishing momentum, the ghost
propagator goes to infinity like $1/p^2$ as the exponent correction $\kappa$ goes to zero
for increasing volume \cite{cuc} and finally, the running coupling is seen to reach no fixed point.

This scenario has been recently derived by us \cite{fra1,fra2} using a reformulation of quantum
field theory in the infrared limit \cite{fra3,fra4}. In order to derive this proper description
of the Yang-Mills theory in the infrared, we have mapped any solution of the Yang-Mills theory
to a $\lambda\phi^4$ theory. This mapping is exact for the classical models while for quantum theories is just
retained in the infrared due to the large quantum corrections that appear in the ultraviolet
\cite{gw,po}.

The aim of this paper is to present a complete mathematical derivation of this scenario.
Some of the main points to be discussed are about the negligible importance of quantum
corrections in the infrared regime. This possibility derives from a relevant result
we obtained recently that a strongly perturbed quantum system behaves semiclassically \cite{fra5}.
This result, when taken to quantum field theory, means that, differently from ultraviolet
computations, an infrared quantum field theory has negligible renormalization effects.
We aim in this way to present a mathematically consistent proof of the form the
gluon and ghost propagators have in the infrared regime.

The paper is structured in the following way. In sec.\ref{sec2} we give a proof of the
equivalence of the classical Yang-Mills and $\lambda\phi^4$ theories. We will show that
scale invariance is retained as we have a scalar massless theory. In sec.\ref{sec3}
we show that the infrared limit of the $\lambda\phi^4$ theory so obtained has
negligible quantum corrections. In sec.\ref{sec4} we derive the propagator
that is the same for the gluon field. After noting that the ghost field decouples
we show that is exactly $\kappa=0$. Consideration on the spectrum are also given.
Finally, in sec.\ref{sec5} we present the conclusions.

\section{Classical equivalence between a massless scalar field and Yang-Mills theories\label{sec2}}

We can state the equivalence between the actions of the scalar field and the Yang-Mills field as: 

\begin{theorem}[Mapping]
\label{teo1}
An extremum of the action
\begin{equation}
    S = \int d^4x\left[\frac{1}{2}(\partial\phi)^2-\frac{\lambda}{4}\phi^4\right]
\end{equation}
is also an extremum of the SU(N) Yang-Mills Lagrangian when 
we properly choose $A_\mu^a$ with some components being zero and all others being equal, and
$\lambda=Ng^2$, being $g$ the coupling constant of the Yang-Mills field.
\end{theorem}

\begin{proof}
We show the existence of the mapping for SU(2) but the proof can be straightforwardly extended to any group given the product between its structure constants. 
The proof of this theorem is straightforwardly obtained by 
a proper
substitution of the
extremum of the scalar field action into the Yang-Mills action. For the latter we take \cite{nair}
\begin{eqnarray}
    S&=&\int d^4x\left[\frac{1}{2}\partial_\mu A^a_\nu\partial^\mu A^{a\nu}
    +\partial^\mu\bar c^a\partial_\mu c^a\right] \\ \nonumber
    &+&\int d^4x\left[gf^{abc}\partial_\mu A_\nu^aA^{b\mu}A^{c\nu}
	+\frac{g^2}{4}f^{abc}f^{ars}A^b_\mu A^c_\nu A^{r\mu}A^{s\nu}
	+gf^{abc}\partial_\mu\bar c^a A^{b\mu}c^c\right].
\end{eqnarray}
where we have taken a coupling with an external field (ghost) $\bar c^a, c^a$ that will turn out to be useful in the following.
In our case one has
\begin{equation}
   f^{abc}f^{ars}=\epsilon_{abc}\epsilon_{ars}=\delta_{br}\delta_{cs}-\delta_{bs}\delta_{cr}
\end{equation}
being $\epsilon_{abc}$ the Levi-Civita symbol. This gives for the quartic term
\begin{equation}
   V(A)=(A_{\mu}^aA^{a\mu})^2-(A_\mu^aA^{a\nu})(A^{b\mu}A^b_\nu).
\end{equation}
Now we adopt Smilga's choice \cite{smi} taking for the components of the Yang-Mills field $A_1^1=A_2^2=A_3^3=\phi$ and all others being zero, One gets immediately
\begin{equation}
   V(A)=6\phi^4=V(\phi)
\end{equation}
and the mapping exists. We notice the factor 6 that compensates for the analogous factor appearing in the kinematic term and generates the proper 't~Hooft coupling. 

With the Smilga's choice, the obtained mapping annihilates also the coupling with the ghost field that in this way is shown to decouple from the Yang-Mills field. For consistency
reasons we are taking, also for the ghost field, all equal components.

Finally, we can write down the mapped action as
\begin{equation}
    S=-3\int d^4x\left[\frac{1}{2}\partial_\mu\phi\partial^\mu\phi 
    +\partial\bar c\partial c\right]
    +3\int d^4x\frac{2g^2}{4}\phi^4.
\end{equation}
\end{proof}

This mapping theorem shows that, at a classical level, if we are able to solve the motion equation of the scalar field, we get immediately a solution of the Yang-Mills field equations. We have also obtained, as a by-product, that
an external field, coupled as the ghost field in the quantum version of the Yang-Mills theory, decouples
becoming free. This latter result does not apply e.g. for a fermion field. This could make interesting 
a formulation of QCD for this solution of the Yang-Mills equations.


Infrared analysis as devised in \cite{fra3,fra4} implies that one knows the exact solution of the Hamiltonian space-independent
equations of the classical theory to build a quantum field theory. For a Yang-Mills theory we know from preceding
well-known research due mostly to Savvidy, Matinyan and others \cite{sav1,sav2,sav3} that the solutions to
Hamilton equations are generically chaotic and so these are completely useless as a vacuum solution
to build a quantum field theory in the infrared limit. There is only one possibility left and this
is the case with all the components of the Yang-Mills field being equal that reduces Hamilton equations to
exactly integrable ones. This approach is a kind of ``replica trick'' as we replicate 
a number of
times a
massless scalar field to obtain an exact solution to classical equations of motion of the Yang-Mills theory.
Indeed, we have recently shown that classical Yang-Mills equations of motion can be exactly solved
producing a massive solution using the exact solution of the classical massless scalar field \cite{frae}
giving an explicit example of the ``mapping theorem'' above. This represents the most relevant
reason for this theorem to be at the foundations of our understanding of Yang-Mills theory. This
can be also seen as a means to apply identically the mechanism to generate mass for a massless
scalar field to the gluon field. 

\section{Infrared limit of a massless scalar field theory \label{sec3}}

The mapping proved in theorem \ref{teo1} at a classical level cannot be kept at a quantum level in the
ultraviolet limit. This result has been known for a long time. This can be traced back to the
computation in this limit of the beta function for Yang-Mills theory and QCD \cite{gw,po} and the
corresponding beta function for scalar field theory \cite{ps}. These computations prove that in the
ultraviolet limit quantum fluctuations spoil the equivalence. The aim of the next theorem is to
show that things go differently in the infrared regime, that is, when we have a coupling formally
going to infinity. In this case the equivalence is retained while the result we obtain is exact
giving a limit path integral for the scalar field theory.

\begin{theorem}[Infrared Limit]
\label{teo2}
In the limit $\lambda\rightarrow\infty$ the Euclidean generating functional integral for the scalar field becomes
\begin{equation}
     Z_0[j]={\cal N}
     \exp\left\{-\int d^4x \left[\frac{1}{2}(\partial_t\phi)^2+\frac{\lambda}{4}\phi^4+j\phi\right]\right\}
\end{equation}
and the field behaves semiclassically.
\end{theorem}

\begin{proof}
We consider the Euclidean generating functional
\begin{equation}
     Z[j]=\int [d\phi]
     \exp\left\{-\int d^4x \left[\frac{1}{2}(\partial\phi)^2+\frac{\lambda}{4}\phi^4+j\phi\right]\right\}.
\end{equation}
The limit $\lambda\rightarrow\infty$ has been analyzed in \cite{kov,pmb,be1,par,be2,be3,coo,be4,svai}. These
authors showed that when the full kinematical term $(\partial\phi)^2$ is taken as a perturbation, a perturbation series is obtained that depends on a regularization parameter $a$ with the interesting limit $a\rightarrow 0$ very singular. The reason for this can be traced back on the fact that some dynamics must be allowed at the leading order
for this limit. To recover a proper generating functional without singularity we rescale imaginary time as
\begin{equation}
    \tau = \sqrt{\lambda}t
\end{equation}
in order to have a dynamics at the same order of the potential. We also note that $j$ is arbitrary and can be redefined
by rescaling it and we are left with the transformed functional
\begin{equation}
\label{eq:f1}
     Z[j]=\int [d\phi]
     \exp\left\{-\sqrt{\lambda}
     \int d^4x\left[
     \frac{1}{2}(\partial_\tau\phi)^2+\frac{1}{4}\phi^4+j\phi
     \right]\right\}
     \exp\left\{-\frac{1}{\sqrt{\lambda}}
     \int d^4x\left[
     \frac{1}{2}(\nabla\phi)^2
     \right]\right\}.
\end{equation}
We can now manage this functional to evaluate quantum fluctuations in the limit $\lambda\rightarrow\infty$. We take
\begin{equation}
\label{eq:dphi}
    \phi = \phi_c + \frac{1}{\sqrt{\lambda}}\delta\phi
\end{equation}
being
\begin{equation}
\label{eq:phic}
    -\partial_\tau^2\phi_c+\phi_c^3=-j
\end{equation}
a solution of the classical equation of motion for the scalar field. We can take eq.(\ref{eq:dphi}) for
the field in view of the functional eq.(\ref{eq:f1}) showing that in the limit $\lambda\rightarrow\infty$
gives rise to the semiclassical approximation for a quartic oscillator that exists as proved in \cite{bow,kle}. 
Higher order corrections identify a gradient expansion. In this way one sees immediately
that quantum fluctuations scale as $1/\sqrt{\lambda}$ becoming increasingly small as the coupling gets larger.
In the same way we get the field $\phi_c$ behaving semiclassically at leading order while higher order
corrections take care of the gradient terms.
\end{proof}

We see that this is in agreement with the result already given in \cite{fra5}. As a by-product we get
the following corollary

\begin{corollary}[Existence]
\label{exist}
Quantum mapping between a scalar field and a Yang-Mills field holds in the infrared limit and the
corresponding perturbation series does exists.
\end{corollary}

\begin{proof}
We have shown in theorem \ref{teo2} that in the infrared limit the perturbation series is just a
semiclassical perturbation series proved to exist having quantum fluctuations going like $O(1/\sqrt{\lambda})$
for increasing $\lambda$. This proves both mapping preservation and existence.
\end{proof}

The importance of this result relies on the fact that whatever conclusion one can draw from quantum field theory in the infrared for the scalar field can be immediately attached to the Yang-Mills field. As in this regime the perturbation theory exists, being just a semiclassical approximation, the common wisdom (this is just a prejudice never seen in the literature) that equal components perturbation theory for the Yang-Mills theory is based on an unstable solution is just false and involve a wrong understanding of the mapping obtained in theorem \ref{teo1}. Whatever conclusion obtained on this mistaken opinion should be discarded as we have mathematical evidence against it while no evidence was ever seen for it.

Finally, some considerations are needed about Lorentz invariance. For our aims it is not needed to do computations having explicit Lorentz invariance but, in the end, this is just a gradient expansion, a common way to do perturbation theory and we are granted to recover Lorentz invariance when computations are finished. Anyhow, it would be interesting to get a formulation of a gradient expansion that is explicitly Lorentz invariant. Finally, we note that, similarly to solitons, any scalar function $f(t)$ can be cast into a form $\tilde f(x,t)$ after a Lorentz transformation.

\section{Gluon and ghost propagators \label{sec4}} 

In this section we turn back to Minkowski space. Then, the following theorem holds for the scalar field

\begin{theorem}[Propagator]
\label{teo3}
In the infrared limit the scalar field propagator is given by
\begin{equation}
\label{eq:sgf}
    G(t)=\Lambda\theta(t)\left(\frac{2}{\lambda}\right)^{\frac{1}{4}}
	{\rm sn}\left[\left(\frac{\lambda}{2}\right)^{\frac{1}{4}}\Lambda t,i\right].
\end{equation}
and $G(-t)$ is the back propagating propagator, being ${\rm sn}$ a Jacobi snoidal elliptical function and $\Lambda$ an integration constant.
\end{theorem}

\begin{proof}
We define a propagator or Green function for a given linear differential equation
\begin{equation}
     L\phi=j
\end{equation}
being $L$ a linear differential operator and $j$ a generic function. A propagator or Green function satisfies the equation
\begin{equation}
     LG=\delta
\end{equation}
being $\delta$ a Dirac distribution. Then, the following identity does hold
\begin{equation}
     \phi=\phi_0+G*j
\end{equation}
being $*$ the convolution operator. In quantum field theory, all the information about the dynamics and spectrum of a given theory can be extracted knowing the corresponding propagator. The above theory can be extended to nonlinear equations. So, let us consider a nonlinear dynamical equation
\begin{equation}
\label{eq:nl}
     \partial_t^2\phi+V'(\phi)=j
\end{equation}
and let us define Green function or propagator a solution of the equation
\begin{equation}
    \partial_t^2G+V'(G)=\delta
\end{equation}
we have shown recently that the following small time series does hold in this case \cite{fra6,fra7}
\begin{equation}
\label{eq:gs}
    \phi(t)=\sum_{n=0}^{+\infty}a_n\int dt'G(t-t')(t-t')^nj(t')
\end{equation}
being $a_n$, with $n\ge 1$ having $a_0=1$, some constants depending on initial conditions and their derivatives and the initial values of $j$ and its derivatives and other parameters entering into eq.(\ref{eq:nl}). From this result we realize that the Green function retains its meaning also for nonlinear equations. We note that the leading term of the series is identical to the solution of the linear case. This permit us to complete our proof.

We note that, in order to apply the above approach one has to make time adimensional. The scalar theory we are considering is massless and scale invariant. This means that our solution will depend on an arbitrary constant having the dimension of a mass that we call $\Lambda$. Then, in the following we will assume that time scales as $t\rightarrow\Lambda t$.

So, our aims can be accomplished by noticing that, for a $\lambda\phi^4$ one has the equation for the Green function
\begin{equation}
    \partial_t^2 G + \lambda G^3=\delta
\end{equation}
having undone the normalization on the time variable that we used to prove theorem \ref{teo2}. This equation can be solved exactly. Firstly, we note that that $G(-t)$ is a solution. This will give the backward propagating Green function. Then, the solution is
\begin{equation}
   G(t)=\theta(t)\left(\frac{2}{\lambda}\right)^{\frac{1}{4}}
	{\rm sn}\left[\left(\frac{\lambda}{2}\right)^{\frac{1}{4}}t,i\right].
\end{equation}
We put this solution into eq.(\ref{eq:gs}) obtaining
\begin{equation}
\label{eq:gs2}
    \phi(t)=\sum_{n=0}^{+\infty}a_n\int dt'\theta(t-t')
    \left(\frac{2}{\lambda}\right)^{\frac{1}{4}}{\rm sn}
    \left[\left(\frac{\lambda}{2}\right)^{\frac{1}{4}}(t-t'),i\right](t-t')^nj(t').
\end{equation}
Being a convolution, we can rewrite it as
\begin{equation}
    \phi(t)=\sum_{n=0}^{+\infty}a_n\int dt'\theta(t')
    \left(\frac{2}{\lambda}\right)^{\frac{1}{4}}{\rm sn}
    \left[\left(\frac{\lambda}{2}\right)^{\frac{1}{4}}t',i\right]t'^nj(t-t').
\end{equation}
and change the integration variable as $t'\rightarrow\left(\frac{\lambda}{2}\right)^{\frac{1}{4}}t'$ giving in the end the series
\begin{equation}
    \phi(t)=\sum_{n=0}^{+\infty}a_n\left(\frac{2}{\lambda}\right)^{\frac{n+2}{4}}
    \int dt'\theta(t_1){\rm sn}(t_1,i)t_1^n j\left(t-\left(\frac{2}{\lambda}\right)^{\frac{1}{4}}t_1\right).
\end{equation}
revealing that this is just an asymptotic series with respect to $\lambda$. 

For a $\lambda\phi^4$ theory we have to kept at least the first order correction. This is obtained with $n=4$ and $a_4=\lambda/20$ \cite{fra7}. Now, we show that in the infrared this correction contributes at most for a  negligible constant. Indeed, for the Jacobi snoidal elliptical function the following identity holds\cite{gr}
\begin{equation}
    {\rm sn}(u,i)=\frac{2\pi}{K(i)}\sum_{n=0}^\infty\frac{(-1)^ne^{-(n+\frac{1}{2})\pi}}{1+e^{-(2n+1)\pi}}
    \sin\left[(2n+1)\frac{\pi u}{2K(i)}\right]
\end{equation}
being $K(i)$ the constant
\begin{equation}
    K(i)=\int_0^{\frac{\pi}{2}}\frac{d\theta}{\sqrt{1+\sin^2\theta}}\approx 1.3111028777.
\end{equation}
Fourier transforming the series (\ref{eq:gs2}) and stopping at the first order one gets
\begin{equation}
    \phi(\omega) = G(\omega)j(\omega)+\frac{\lambda}{20}\left[\frac{d^4}{d\omega^4}G(\omega)\right]j(\omega)+\ldots
\end{equation}
with the transformed propagator
\begin{equation}
\label{eq:prop}
    G(\omega)=\sum_{n=0}^\infty\frac{B_n}{\omega^2-\omega_n^2+i\epsilon}
\end{equation}
being
\begin{equation}
    B_n=(2n+1)\frac{\pi^2}{K^2(i)}\frac{(-1)^{n+1}e^{-(n+\frac{1}{2})\pi}}{1+e^{-(2n+1)\pi}},
\end{equation}
and, after undoing time scaling with the arbitrary constant $\Lambda$,
\begin{equation}
\label{eq:ms}
    \omega_n = \left(n+\frac{1}{2}\right)\frac{\pi}{K(i)}\left(\frac{\lambda}{2}\right)^{\frac{1}{4}}\Lambda.
\end{equation}
In the infrared limit, $\omega\rightarrow 0$, we can neglect all the corrections and the leading order term is the most important. This proves the theorem.
\end{proof}

From this theorem we get the following important corollary

\begin{corollary}[Mass Spectrum] 
The mass spectrum of a $\lambda\phi^4$ theory in the infrared is given by
\begin{equation}
    \omega_n = \left(n+\frac{1}{2}\right)\frac{\pi}{K(i)}\left(\frac{\lambda}{2}\right)^{\frac{1}{4}}\Lambda.
\end{equation}
\end{corollary}

We obtained this directly in the proof ot theorem \ref{teo3}. We recognize the spectrum of an harmonic oscillator. This is proper to theories becoming free in some sense. Indeed, through this result one can attach a definition of ``triviality'' to this kind of theories that show a running coupling not reaching a fixed point in the infrared through a renormalization group analysis using small perturbation theory\cite{fra8}.

What is presently emerging from lattice computations at larger volumes for Yang-Mills theory in the infrared is exactly the scenario we have depicted till now for the scalar field theory. This should be expected on the basis of corollary \ref{exist} obtained through theorems \ref{teo1} and \ref{teo2}. So, the following theorem holds

\begin{theorem}[Final]
\label{teo4}
In the infrared limit the gluon propagator is given by
\begin{equation}
    F(\omega,0)=\sum_{n=0}^\infty
    (2n+1)\frac{\pi^2}{K^2(i)}\frac{(-1)^{n+1}e^{-(n+\frac{1}{2})\pi}}{1+e^{-(2n+1)\pi}}
    \frac{1}{\omega^2-\omega_n^2+i\epsilon}
\end{equation}
with a mass spectrum
\begin{equation}
    \omega_n = \left(n+\frac{1}{2}\right)\frac{\pi}{K(i)}\left(\frac{Ng^2}{2}\right)^{\frac{1}{4}}\Lambda.
\end{equation}
The ghost propagator is given by
\begin{equation}
    G(\omega,{\bf k})=\frac{1}{(|{\bf k}|^2-\omega^2)^{1+\kappa}-i\epsilon}
\end{equation}
with $\kappa=0$.
\end{theorem}

\begin{proof}
Corollary \ref{exist} permits to conclude that in the infrared both scalar field theory and Yang-Mills theory share the same propagator as obtained from theorem \ref{teo3} and the same mass spectrum with the substitution $\lambda\rightarrow Ng^2$. We recognize here 't Hooft scaling for Yang-Mills theory. From theorem \ref{teo1} and corollary \ref{exist} we know that the ghost field decouples in the infrared giving the propagator of a free particle. This permits us to conclude that $\kappa=0$.
\end{proof}

From this theorem is easy to see that the gluon propagator is finite in the infrared limit $\omega\rightarrow 0$ and one has a complete agreement with the most recent lattice computations at large volumes.

Finally, we can define a running coupling as \cite{as}
\begin{equation}
   \alpha(\omega)=D(\omega,0)Z^2(\omega,0)
\end{equation}
being
\begin{equation}
   F(\omega,0)=\frac{D(\omega,0)}{\omega^2}
\end{equation}
and
\begin{equation}
   G(\omega,0)=-\frac{Z(\omega,0)}{\omega^2}
\end{equation}
then one has immediately that
\begin{equation}
   \lim_{\omega\rightarrow 0}\alpha(\omega) = 0
\end{equation}
in complete agreement with lattice results. 

\section{Conclusions\label{sec5}}

Lattice computations can probe furhter this scenario by an in-depth analysis of both scalar field and Yang-Mills theories by comparing the evaluation of the fields and, most importantly, deriving the mass spectrum in the infrared for the scalar field. But the present support by recent computations to these results is already impressive. This makes real peculiar and somewhat unexpected the behavior of the free Yang-Mills theory in the infrared. We see that it develops a mass gap due to the strong effect of the self-interaction originating by the non-linear part of the dynamics. On the other side, the theory has a running coupling that does not reach any fixed point in the infrared contrarily to a naive expectation. This entails some sort of triviality, the same seen for the scalar field that manifests itself through the simple spectrum of an harmonic oscillator. While we presently see strong agreement for the spectrum in the Yang-Mills theory as proved by comparing our mathematical analysis and lattice computations \cite{fra2,fra4}, we would like to see a similar effort to be spent for the scalar field in four dimensions, the only case where the mapping seems to be successfully accomplished due to the fact that couplings are pure numbers in this case.

Extensions to this mathematical analysis could be realized by obtaining the possibility to study scattering in the infrared by identifying proper asymptotic states and to extend the analysis with the introduction of quarks analyzing the hadron spectrum that in this way is obtained. Anyhow, the road is open for an intensive study of quantum field theories with a large coupling, a regime not accessible until now through perturbative methods.

\begin{acknowledgments}
I would like to thank Ernst-Michael Ilgenfritz for the very useful correspondence and for sending
me a copy of posters he and his group presented at Lattice 2007.
\end{acknowledgments}


\begin{thebibliography}{99}
\bibitem{as}R. Alkofer, L. von Smekal, Phys. Rept. {\bf 353}, 281 (2001).
\bibitem{cf} C. S. Fischer, J.Phys. G {\bf 32},  R253-R291 (2006).
\bibitem{cf2} C. S. Fischer, A. Maas, J. M. Pawlowski, L. von Smekal, Ann. Phys. {\bf 322}, 2916 (2007). 
\bibitem{ilg1} I.L. Bogolubsky, E.–M. Ilgenfritz, M. M\"uller-Preussker, A.Sternbeck, PoS (LATTICE 2007) 290.
\bibitem{ilg2} I.L. Bogolubsky, V.G. Bornyakov, G. Burgio, E.–M. Ilgenfritz,
V.K. Mitrjushkin, M. M\"uller-Preussker, P. Schemel, PoS (LATTICE 2007) 318.
\bibitem{cuc} A. Cucchieri, T. Mendes, PoS (LATTICE 2007) 297.
\bibitem{ste} A. Sternbeck, L. von Smekal, D. B. Leinweber, A. G. Williams, PoS (LATTICE 2007) 340.
\bibitem{fra1} M. Frasca, Int. J. Mod. Phys. A {\bf 22}, 2433 (2007).
\bibitem{fra2} M. Frasca, arXiv:0704.3260 [hep-th].
\bibitem{fra3} M. Frasca, Phys. Rev. D {\bf 73}, 027701 (2006); Erratum-ibid., 049902 (2006).
\bibitem{fra4} M. Frasca, Int. J. Mod. Phys. A {\bf 22}, 1727 (2007).
\bibitem{sav1} S. G. Matinyan, G. K. Savvidy, N. G. Ter-Arutunian Savvidy, Sov. Phys. JETP {\bf 53}, 421 (1981).
\bibitem{sav2} G. K. Savvidy, Phys. Lett. B {\bf 130}, 303 (1983).
\bibitem{sav3} G. K. Savvidy, Nucl. Phys. B {\bf 246}, 302 (1984).
\bibitem{frae} M. Frasca, arXiv:0807.2179 [hep-th].
\bibitem{gw} D. J. Gross, F. Wilczek, Phys. Rev. Lett. {\bf 30}, 1343 (1973).
\bibitem{po} H. D. Politzer, Phys. Rev. Lett. {\bf 30}, 1346 (1973).
\bibitem{fra5} M. Frasca, Proc. R. Soc. A {\bf 463}, 2195 (2007).
\bibitem{nair} V. P. Nair, {\sl Quantum Field Theory}, (Springer, New York, 2005), p.194.
\bibitem{smi} A. Smilga, {\sl Lectures on Quantum Chromodynamics}, 
(World Scientific, Singapore, 2001).
\bibitem{ps} M. E. Peskin, D. V. Schroeder, {\sl An Introduction to Quantum Field Theory},
(Perseus, Reading, 1995), Ch.12.
\bibitem{kov} S. Kovesi-Domokos, Nuovo Cim. A {\bf 33}, 769 (1976).
\bibitem{pmb} R. Benzi, G. Martinelli, and G. Parisi, Nucl. Phys. B {\bf 135}, 429 (1978).
\bibitem{be1} C. M. Bender, F. Cooper, G. S. Guralnik, and D. H. Sharp, Phys. Rev. D {\bf 19}, 1865 (1979).
\bibitem{par} N. Parga, D. Toussaint, and J. R. Fulco, Phys. Rev. D {\bf 20}, 887 (1979).
\bibitem{be2} C. M. Bender, F. Cooper, G. S. Guralnik, D. H. Sharp, R. Roskies, and 
M. L. Silverstein, Phys. Rev. D {\bf 20}, 1374 (1979).
\bibitem{be3} C. M. Bender, F. Cooper, G. S. Guralnik,
H. Moreno, R. Roskies, and D. H. Sharp, Phys. Rev. Lett. {\bf 45}, 501 (1980).
\bibitem{coo} F. Cooper, and R. Kenway, Phys. Rev. D {\bf 24}, 2706 (1981).
\bibitem{be4} C. Bender, F. Cooper, R. Kenway, and L. M. Simmons, Phys. Rev. D
{\bf 24}, 2693 (1981).
\bibitem{svai} N. F. Svaiter, Physica A {\bf 345}, 517 (2005).
\bibitem{bow} C. M. Bender, K. Olaussen and P. S. Wang, Phys. Rev. D {\bf 16}, 1740 (1977).
\bibitem{kle} H. Kleinert, {\sl Path Integrals in Quantum Mechanics, Statistics, and Polymer Physics, and Financial Markets}, (World Scientific Publishing, Singapore, 2004).
\bibitem{fra6} M. Frasca, Mod. Phys. Lett. A {\bf 22}, 1293 (2007). 
\bibitem{fra7} M. Frasca, Int. J. Mod. Phys. A {\bf 23}, 299 (2008).
\bibitem{gr} I. S. Gradshteyn, I. M. Ryzhik, {\sl Table of Integrals, Series, and Products},
(Academic Press, 2000). 
\bibitem{fra8} M. Frasca, Int. J. Mod. Phys. A {\bf 22}, 2433 (2007).
\end{thebibliography}
\end{document}